\def\imat {i}

\documentclass[prb,twocolumn,showpacs,aps]{revtex4}
\usepackage{graphicx}

\begin{document}

\title{Continuous quantum feedback of coherent oscillations in a solid-state
qubit}

\author{Qin Zhang, Rusko Ruskov,\footnote{On leave of absence from Institute
for Nuclear Research and Nuclear Energy, Sofia BG-1784, Bulgaria;
Present address:
Physics Department, Pennsylvania State University, University Park, PA 16802}
and Alexander N. Korotkov\footnote{Electronic mail: korotkov@ee.ucr.edu}}
\affiliation{
Department of Electrical Engineering, University of California,
Riverside, CA 92521-0204.
}
\date{\today}

\begin{abstract}
    We have analyzed theoretically the operation of the Bayesian quantum
feedback of a solid-state qubit, designed to maintain perfect coherent 
oscillations in the qubit for arbitrarily long time. In particular,
we have studied the feedback efficiency in presence of dephasing environment 
and detector nonideality. Also, we have analyzed the effect of qubit
parameter deviations and studied the quantum feedback control of an 
energy-asymmetric qubit.
\end{abstract}
%\pacs{PACS numbers: }
\pacs{73.23.-b; 03.65.Ta; 03.67.Lx}
%73.23.-b Electronic transport in mesoscopic systems
%03.65.Ta Foundations of Q. Mech; measurement theory
%03.67.Lx Quantum computation 
%85.35.-p Nanoelectronic devices

\maketitle

%\newpage
%\narrowtext

%\vspace{0.6cm}

\section{Introduction}

    Continuous quantum feedback in optics and atomic physics has been
studied theoretically \cite{Wiseman,Tombesi,Hofmann,Wang,Doherty}
for more than a decade
(see also Refs.\ \onlinecite{Shapiro,Yamamoto,Caves-87,Wiseman-94,Doherty-3})
and has been recently demonstrated
experimentally.\cite{Geremia} In contrast, continuous quantum feedback
in solid-state mesoscopics is a relatively new subject.
\cite{Ruskov-02,Ruskov-SPIE,Korotkov-04,Hopkins,Ruskov-squeez} 
The use of quantum feedback to maintain coherent (Rabi) oscillations in a
qubit for arbitrarily long time has been proposed and analyzed in Refs.\
\onlinecite{Ruskov-02} and \onlinecite{Ruskov-SPIE}; a simplified experiment
has been proposed in Ref.\ \onlinecite{Korotkov-04}. Cooling of
a nanoresonator by quantum feedback has been proposed and analyzed in Ref.\
\onlinecite{Hopkins}. The use of quantum feedback for the nanoresonator
squeezing has been studied in Ref.\ \onlinecite{Ruskov-squeez}.

    Feedback control of a quantum system requires continuous monitoring
of its evolution (in ideal case the wavefunction should be monitored),
which is the main non-trivial feature of quantum feedback.
 Obviously, the operation of quantum feedback cannot be analyzed using
the ``orthodox'' approach of instantaneous collapse, \cite{Neumann}
which is not suitable for continuous quantum measurement. Also, the
ensemble-averaged (``conventional'') approach \cite{Caldeira}
to continuous quantum measurement is not suitable since it cannot
describe random evolution of a single quantum system.
Therefore, analysis of quantum feedback requires a special theory
capable of describing continuous measurement of a single quantum system.

    The development of such theories has started long ago
\cite{Mensky,Gisin,Caves-86} and has attracted most of attention in
quantum optics \cite{Wiseman,Carmichael,Plenio}
(in spite of similar underlying principles, the theories may differ
significantly in formalism and area of application).
For solid-state qubits such theory (``Bayesian'' formalism) has been
developed relatively recently \cite{Kor-99,Kor-01} (for review see Ref.\
\onlinecite{Kor-rev}). The equivalence of the Bayesian formalism
to the quantum trajectory approach translated \cite{Goan,Goan-2,Oxtoby}
from quantum optics has been shown in Ref.\ \onlinecite{Goan}.

    In simple words, the Bayesian formalism takes into
account the information contained in the noisy output of the solid-state
detector measuring the qubit, so that the corresponding quantum back-action
onto qubit evolution is described explicitly. In classical probability
theory the way to deal with an incomplete information is via the Bayes
formula;\cite{Feller} 
 it can be shown \cite{Kor-99,Kor-01,Kor-rev} that a somewhat similar
procedure should be used for evolution of the qubit density matrix
due to continuous measurement, that explains why the formalism is called
Bayesian. The Bayesian formalism shows that an ideal (quantum-limited)
detector can monitor precisely the random evolution of the qubit wavefunction
in the course of measurement; and if the measurement starts with a mixed
state, the qubit density matrix is gradually purified, eventually
approaching a pure state. The quantum point contact (QPC) is an example
of (theoretically) ideal detector. When the detector does not have 100\%
quantum efficiency [as in the case of a single-electron transistor (SET)],
there is an extra dephasing term in the evolution equation, so that
the qubit purification due to gradually acquired information competes
with the decoherence due to detector nonideality.

    The possibility to monitor the random quantum evolution of the qubit
in the process of measurement naturally allows us to arrange a feedback
loop which keeps the qubit evolution close to a desired ``trajectory''.
Of course, the measurement process disturbs the qubit evolution; however,
the detector output contains enough information to monitor and undo
the effect of this disturbance.
It is important that the deviations from the desired trajectory due to
interaction with decohering environment are efficiently suppressed
by the feedback loop, that can be useful, for example, in a quantum
computer. The feedback loop considered in Refs.\ \onlinecite{Kor-01} and 
\onlinecite{Ruskov-02} has been designed to maintain the coherent 
oscillations in the qubit for arbitrarily long time by comparing the 
oscillation phase with the desired value and keeping the phase difference 
close to zero (the amplitude of oscillations is equal to unity in the case 
of ideal detector and energy-symmetric qubit). 
It has been shown that the fidelity of such feedback loop 
can be arbitrarily close to 100\%, while it decreases in the case of a
nonideal detector and/or significant interaction with environment as well
as in the case of finite bandwidth of the line carrying the signal
from detector.

    The present paper is a more detailed analysis of the operation
of the feedback loop proposed in Refs.\ \onlinecite{Kor-01} and
\onlinecite{Ruskov-02}. In particular, 
we study the feedback loop operation in presence of extra dephasing 
due to environment and nonideal detector, analyze the effect of 
qubit parameter deviation, and consider the feedback of a qubit
with energy asymmetry.
In the next Section we describe the model, in Section III we consider 
the feedback operation in the ideal case, Section IV is devoted to 
the effects of nonideal detector and extra dephasing, in Section V 
we analyze the worsening of feedback efficiency in the case of qubit
parameter deviations, in Section VI we study the feedback of 
an energy-asymmetric qubit, and Section VII is a conclusion.

\section{Model}

\begin{figure}
\centering
\includegraphics[width=2.8in]{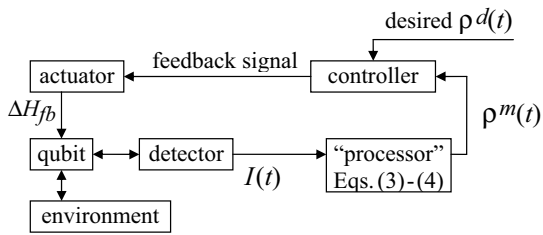}
%\vspace{0.4cm}
\caption{ Schematic of the quantum feedback loop maintaining the 
quantum oscillations in a qubit.
The qubit oscillations affect the current $I(t)$ through a weakly coupled
detector; this signal is translated by the ``processor'' into continuously 
monitored value $\rho^m (t)$ of the qubit density matrix. Next, by comparing
$\rho^m(t)$ with the desired oscillating state $\rho^d(t)$, a
certain algorithm (``controller'') produces the feedback signal applied
to an ``actuator'' which changes the qubit tunneling amplitude $H+\Delta 
H_{fb}$, in order to reduce the difference between $\rho^m$ and $\rho^d$. 
 }
\label{schematic}\end{figure}

    Let us consider the quantum feedback loop shown in Fig.\
\ref{schematic}, which controls the qubit characterized by the Hamiltonian
    \begin{equation}
{\cal H}_{qb} = \frac{\varepsilon_{fb}}{2}\,
(c_2^\dagger  c_2 - c_1^\dagger c_1)
+ H_{fb} (c_1^\dagger c_2 +c_2^\dagger c_1),
    \label{H_qb}
    \end{equation}
where $c_{1,2}^\dagger$ and $c_{1,2}$ are creation and annihilation operators
corresponding to two ``localized'' states of the qubit, representing the
``measurement basis''. The qubit energy asymmetry $\varepsilon_{fb}$ and
tunneling amplitude $H_{fb}$ can both be controlled by the feedback loop:
    \begin{equation}
H_{fb} = H+ \Delta H_{fb}, \hspace{0.5cm}
\varepsilon_{fb}= \varepsilon + \Delta\varepsilon_{fb};
    \end{equation}
however, in this paper we assume $\Delta \varepsilon_{fb}=0$, 
so that only tunneling is controlled. Intrinsic frequency of
coherent oscillations in the qubit (without interaction and feedback)
is $\Omega =\sqrt{4H^2+\varepsilon^2}/\hbar$; we also call it Rabi frequency,
not implying presence of microwave radiation (despite this terminology
differs from the initial meaning of Rabi oscillations, it is conventionally 
used nowadays).

    For definiteness we consider a ``charge'' qubit continuously
measured by QPC or SET, so that the measurement setup is similar to what
has been studied theoretically, e.g. in Refs.\
\onlinecite{Gurvitz,Shnirman-98,Kor-99,Kor-01,Kor-rev,Averin,Goan,Goan-2,%
Oxtoby,Buttiker,Clerk}. 
Taking into account the quantum back-action due to measurement,
the evolution of the qubit density matrix $\rho$ is described by
the Bayesian equations \cite{Kor-99,Kor-01,Kor-rev} (in Stratonovich form)
        \begin{eqnarray}
&& \hspace{-0.7cm} \dot{\rho}_{11}=
-2\,\frac{H_{fb}}{\hbar}\,\mbox{Im}\rho_{12}
         +\rho_{11}\rho_{22} \, \frac{2\Delta I}{S_I} \, [I(t)-I_0],
        \label{Bayes1}\\
&& \hspace{-0.7cm}
{\dot\rho}_{12}=  \imat \, \frac{\varepsilon_{fb}}{\hbar }\,\rho_{12}+
        \imat \, \frac{H_{fb}}{\hbar } \, (\rho_{11}-\rho_{22})
        \nonumber \\
&&  -( \rho_{11}-  \rho_{22})  \frac{\Delta I}{S_I} \,
[I(t)-I_0]\, \rho_{12} -\gamma \rho_{12} \, ,
        \label{Bayes2}
        \end{eqnarray}
where $I(t)$ is the noisy detector current (output signal),
$S_I$ is the spectral density of current noise,
$\Delta I=I_1-I_2$ is the difference between two average currents
$I_1$ and $I_2$ corresponding to the two qubit states, and
$I_0=(I_1+I_2)/2$. The dephasing rate $\gamma =\gamma_d+\gamma_{env}$
has the contribution $\gamma_d$ due to detector nonideality,
$\gamma_d= (\eta^{-1}-1)(\Delta I)^2/4S_I$
(here $\eta \leq 1$ is the quantum efficiency 
\cite{Kor-99,Kor-01,Kor-rev,Averin,Buttiker,Clerk})
and contribution
$\gamma_{env}$ due to interaction with extra environment.
As always, $\rho_{11}+\rho_{22}=1$ and $\rho_{21}=\rho_{12}^*$.
Equations (\ref{Bayes1})--(\ref{Bayes2}) imply weak detector response
$|\Delta I| \ll I_0$, quasicontinuous current, and large detector voltage
compared to the qubit energy. The current
                \begin{equation}
I(t) = I_0 + ( \Delta I/2) (\rho_{11} -\rho_{22})  +\xi (t)
        \label{I(t)}
    \end{equation}
has the pure noise contribution $\xi (t)$ with frequency-independent
spectral density $S_I$. Notice that averaging of Eqs.\
(\ref{Bayes1})--(\ref{Bayes2}) over $\xi (t)$ leads to the standard
ensemble-averaged equations \cite{Caldeira} with ensemble dephasing
rate $\Gamma =(\Delta I)^2/4S_I +\gamma$.
We characterize coupling between qubit and detector by the
dimensionless constant ${\cal C}=\hbar (\Delta I)^2/S_I H$
(we assume\cite{H-sign} $H>0$) and concentrate 
on the case of weak coupling ${\cal C}\alt 1$ (notice that ${\cal C}=1$
can still be considered a weak coupling since the quality factor of
oscillations in presence of measurement \cite{Kor-sp} is $8\eta/{\cal C}$
for $\varepsilon =0$).

    In this paper we consider the ``Bayesian'' feedback, \cite{Ruskov-02}
which requires a ``processor'' solving Eqs.\ (\ref{Bayes1})--(\ref{Bayes2})
in real time -- see Fig.\ \ref{schematic} (other possibilities are,
for example, ``direct'' feedback briefly mentioned in Ref.\
\onlinecite{Ruskov-02} and ``simple'' feedback via quadrature components
analyzed in Ref.\ \onlinecite{Korotkov-04}). 
In this paper we neglect the effect of finite bandwidth 
\cite{Ruskov-02,Ruskov-SPIE,Oxtoby-2} of the line carrying the detector 
signal, and we also neglect the signal delay in the feedback loop. 
As a result, in most of the paper we assume that the monitored value 
$\rho^m(t)$ of the qubit density matrix coincides with the actual 
value $\rho (t)$. Only in Section V we consider $\rho^m$ different from 
$\rho$  because of the deviation of the qubit parameters 
$H$ and $\varepsilon$ from the values assumed in the ``processor'' 
(finite signal bandwidth would also lead to difference between $\rho$ and
$\rho^m$).

    For the feedback control the monitored qubit evolution
is compared with the desired evolution (Fig.\ \ref{schematic}),
and the difference signal is used to control the qubit parameters
in order to decrease the difference.
Actually, various algorithms (``controllers'') are possible
for this purpose; in this paper we will consider linear control
(see below). We study the feedback loop, which goal is to maintain
perfect coherent oscillations in the qubit for arbitrarily long time, 
and (except for Section VI) the desired evolution is 
    \begin{equation}
\rho_{11}^d= \frac{1+\cos \Omega_0 t}{2}, \hspace{0.3cm}
\rho_{12}^d=i\, \frac{\sin \Omega_0 t}{2},
        \label{rho-desired}
    \end{equation}
with frequency $\Omega_0=2H/\hbar$ corresponding to $\varepsilon =0$.
Except for Section VI, we assume the following feedback control: 
    \begin{eqnarray}
 &&  \hspace{-0.7cm}  \Delta H_{fb} = -F H \Delta \phi_m , 
    \label{dHfb} \\
 && \hspace{-0.7cm} \Delta \phi_m = \phi_m (t) -\Omega_0 t \,\,\, 
(\mbox{mod}\, 2\pi),
     \\ 
&& \hspace{-0.7cm} \phi_m (t) = \mbox{arctan} [2\,\mbox{Im}
\rho_{12}^m / (\rho_{11}^m -\rho_{22}^m )] \,\,
        \nonumber \\
&&\hspace {0.5cm} + (\pi /2) [1-\mbox{sgn}(\rho_{11}^m -\rho_{22}^m )] 
%\,\, \,\, (\mbox{mod}\,2\pi)
,
        \label{phi_m}
    \end{eqnarray}
where $\phi_m$ is the monitored value of the phase, phase
difference $\Delta \phi_m $ is defined as $|\Delta \phi_m|\leq \pi$,
and $F$ is a dimensionless feedback factor [the second term in Eq.\
(\ref{phi_m}) provides proper phase continuity on $2\pi$ circle]. 
The controller (\ref{dHfb}) is  supposed to decrease the phase difference
(negative feedback): if phase $\phi_m(t)$ is ahead of the desired
value, then $\Delta H_{fb}$ is negative, that slows down the qubit
oscillations and decreases the phase shift; if $\phi_m(t)$ is
behind the desired value, the oscillation frequency increases to
catch up.

    We will characterize the feedback efficiency by the
``synchronization degree'' $D$ defined as averaged over time scalar product
of two Bloch vectors corresponding to the desired and actual 
states of the qubit. An equivalent definition is
    \begin{equation}
    D=2\langle \mbox{Tr} \rho \rho^d \rangle -1,
     \label{D-def}
    \end{equation}
where $\langle .. \rangle$ denotes averaging over time. Perfect feedback
operation corresponds to $D=1$ (notice that $\rho^d$ is a pure state).
Feedback efficiency $D$ can be easily translated into average fidelity as
$(D+1)/2$ or $\sqrt{(D+1)/2}$, depending on the definition of
fidelity \cite{Jozsa,Doherty-3} (translation formula would be slightly 
longer if neither $\rho$ nor $\rho^d$ are pure states). 
 We prefer to use $D$ instead of fidelity because
$D=0$ in absence of feedback when $\rho$ and $\rho^d$ are completely
uncorrelated, while fidelity is non-zero.

    \section{Ideal case}

    Let us start analysis with the basic ideal case of $\eta =1$ 
(quantum-limited detector, e.g.\ QPC), absence of extra environment
($\gamma_{env}=0$), and symmetric qubit 
($\varepsilon =0$). The analytical results for this case have been
presented in Ref.\ \onlinecite{Ruskov-02}; here we discuss the
derivation in more detail.

    Since $\eta =1$ and $\gamma_{env} =0$, so that there is no dephasing
term in Eq.\ (\ref{Bayes2}), the qubit density matrix $\rho$ becomes pure
in the process of measurement. \cite{Kor-rev} 
Because of the energy symmetry, $\varepsilon_{fb}=\varepsilon =0$, 
the real part of $\rho_{12}$ eventually becomes zero. This happens because
the product $(\rho_{11}-\rho_{22})(I-I_0)$ affecting the evolution
of $\mbox{Re} \rho_{12}$ in Eq.\ (\ref{Bayes2}) is on average positive. 
Therefore, after a transient period the evolution of the density matrix 
$\rho$ can be parameterized as 
        \begin{equation}
\rho_{11}= (1+\cos \phi )/2, \,\,\,\, \rho_{12} =i(\sin \phi)/2
        \label{phi-def}
        \end{equation}
 with only one
parameter $\phi (t)$. We have also checked this fact numerically.
Notice that since the qubit is monitored exactly, 
$\rho =\rho^m$, the phase $\phi$ coincides (modulo $2\pi$) with the 
monitored phase $\phi_m$ defined by Eq.\ (\ref{phi_m}). 

        The evolution equation for phase $\phi$ can be easily derived from
Eq.\ (\ref{Bayes2}) as 
        \begin{equation}
\dot\phi = 2H_{fb}/\hbar - (\Delta I/S_I)(I-I_0)\sin \phi ,
        \end{equation}
so the phase difference $\Delta \phi =\phi -\Omega_0 t$ (which 
coincides with $\Delta \phi_m$) evolves as 
    \begin{equation}
\frac{d}{dt}\, \Delta \phi = -\sin \phi \, \frac{\Delta I}{S_I}
\left( \frac{\Delta I}{2}\, \cos\phi +\xi \right) -
F\Omega_0\,\Delta\phi .
        \label{ddphi}
    \end{equation}
(All equations are in the Stratonovich form, so we use usual rules
for derivatives.\cite{Oksendal}) Notice that because of our definition 
$|\Delta \phi |\leq \pi$, the phase difference jumps by $\pm 2\pi$
at the borders of $\pm \pi$ interval. 

        For weak coupling (${\cal C}/8\ll 1$) the qubit oscillations 
are only slightly perturbed by measurement and corresponding phase
diffusion is relatively slow. Assuming that the feedback control 
is also slow on the timescale of oscillations ($|\Delta H_{fb}| \ll H$),
we can average Eq.\ (\ref{ddphi}) over relatively fast oscillations.
Then the first term in parentheses is averaged to zero and averaging
of the term $-(\sin \phi)(\Delta I/S_I)\xi (t)$ leads to the effective 
noise $\tilde\xi (t)$ with spectral density $S_{\tilde\xi}=(\Delta I)^2/2S_I$,
so that the remaining slow evolution of phase difference is
    \begin{equation}
\frac{d}{dt}\, \Delta \phi = \tilde{\xi}
- F\Omega_0\,\Delta\phi . 
        \label{ddphi2}
    \end{equation}

        To find the feedback efficiency $D=\langle \cos \Delta \phi \rangle$
analytically, let us also assume that feedback performance is good enough 
to keep
the phase difference $\Delta \phi$ well inside the $\pm \pi$ interval, 
so that the phase slips (jumps of $\Delta \phi$ by $\pm 2\pi$) occur 
sufficiently rare. In this case we can consider Eq.\ (\ref{ddphi2}) 
on the infinite interval of $\Delta \phi$. The corresponding 
Fokker-Planck-Kolmogorov 
equation for the probability density $\sigma (\Delta \phi )$
        \begin{equation}
\frac{\partial \sigma }{\partial t}  = \frac{\partial}{\partial \Delta \phi} 
\left( \sigma \, F\Omega_0\, \Delta \phi \right) 
+\frac{1}{4}\, \frac{\partial^2 (S_{\tilde\xi} \sigma )}
{\partial (\Delta \phi)^2} 
        \label{F-P}
        \end{equation}
has the Gaussian stationary solution 
$\sigma_{st}(\Delta \phi) = (2\pi {\cal V} )^{-1/2}\linebreak[1] \exp 
[-(\Delta \phi)^2 /2 {\cal V}]$ with variance 
${\cal V}=S_{\tilde\xi}/4F\Omega_0 = {\cal C}/16F$. 
Therefore, $\langle \cos \Delta \phi \rangle = \linebreak[1] 
\exp (-{\cal V}/2)$,
and so the feedback efficiency is \cite{Ruskov-02} 
        \begin{equation}
D=\exp (-{\cal C}/32 F)
        \label{D-analyt}
        \end{equation}
in the case of weak coupling and sufficiently efficient feedback
(${\cal C}\alt 1$, $D\agt 1/2$). 

\begin{figure}
\centering
\includegraphics[width=3.0in]{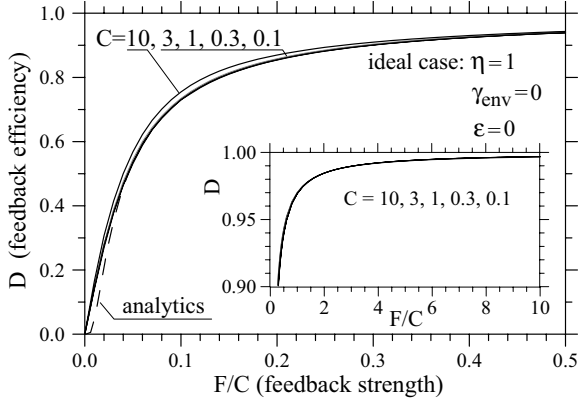} 
%\vspace{0.4cm}
\caption{Solid lines: quantum feedback efficiency $D$ as a function 
of the feedback strength $F$ for five different values of the coupling 
${\cal C}$ and ideal operation conditions (see text).
 The curves for ${\cal C}\leq 3$ practically coincide with each other. 
Dashed line shows the analytical result (\ref{D-analyt}). Inset shows
the same curves for larger range of $F/{\cal C}$.
 }
\label{ideal}\end{figure}

        Figure \ref{ideal} shows comparison between the analytical result
(\ref{D-analyt}) and numerical results for $D$ as a function 
of feedback factor $F$ (scaled by coupling ${\cal C}$). Numerical results
have been obtained by direct simulation of the Bayesian equations 
(\ref{Bayes1})--(\ref{Bayes2}) using the Monte Carlo method 
\cite{Kor-99,Kor-01} for five values of coupling: ${\cal C}=10$, 3, 1, 0.3, 
and 0.1. 
 One can see that for weak coupling ${\cal C}\alt 1$ the analytics works 
very well when the feedback is sufficiently efficient, $D\agt 0.5$. 
Another important observation is that with the feedback factor $F$ normalized
by coupling ${\cal C}$, the curves for ${\cal C} \leq 1$ are practically
indistinguishable from each other, the curve for ${\cal C}=3$ goes
a little higher but still within the line thickness, and only the curve
for ${\cal C}=10$ is noticeably different. Therefore, as expected, 
the weak-coupling limit is practically reached starting with 
${\cal C} \leq 1$. This makes unnecessary to analyze numerically the case
of very small coupling ${\cal C}\ll 1$, which requires much longer
simulation time than the case of moderately small coupling.

Notice that $|\Delta H_{fb}|/H <\pi F$, and $F$ scales with coupling 
${\cal C}$. Therefore, in the experimentally realistic case ${\cal C}\ll 1$ 
a typical amount of the parameter change due to feedback is small, 
$|\Delta H_{fb}| \ll H$. [Hence, we should not worry about unnatural 
assumption of using control equation (\ref{dHfb}) even when $H_{fb}$
becomes negative.] 

   The feedback efficiency $D$ is directly related \cite{Korotkov-04} 
to the average in-phase quadrature component of the detector current, 
$\langle I(t) \cos \Omega_0 t\rangle =(\Delta I/4 )[D+\langle \cos 
( 2\Omega_0 t+\Delta \phi )\rangle ]$, so that in the case of practically 
harmonic oscillations $D= (4/\Delta I) \langle I(t) \cos \Omega_0 t\rangle $.
Positive in-phase quadrature is one of easy ways to verify the quantum
feedback operation experimentally.

        Besides analyzing feedback efficiency $D$, let us also calculate
the qubit correlation function $K_z(\tau)=\langle z(t+\tau) \, z(t)\rangle$ 
where $z=\rho_{11}-\rho_{22}$. In the case of practically harmonic
(weakly disturbed) oscillations, the correlation function 
$K_z(\tau )=\langle \cos [\phi (t+\tau )]\cos [\phi (t)] \rangle$ is 
equal to $\langle \cos [\Omega_0 \tau +\delta \phi (\tau )] \rangle /2$
where $\delta \phi (\tau )=\Delta \phi (t+ \tau) -\Delta \phi (t)$
is the phase deviation during time $\tau$. Since in our case 
$\langle \sin \delta \phi (\tau )\rangle =0$ because of the symmetry of 
Eq.\ (\ref{ddphi2}), the correlation function is reduced to
        \begin{equation}
K_z(\tau )= \frac{\cos \Omega_0\tau }{2}\, \langle \cos \delta \phi (\tau )
\rangle . 
        \label{K_z-via-dphi}
        \end{equation}
  We can find  $\langle \cos \delta \phi (\tau )\rangle $ using exact
solution of the Fokker-Planck-Kolmogorov equation (\ref{F-P}) with
initial condition 
$\sigma (\Delta \phi, 0)=\delta (\Delta \phi-\Delta\phi_0)$: 
        \begin{eqnarray} 
&& \hspace{-0.8cm}
 \sigma (\Delta \phi, \tau |\Delta\phi_0) = \frac{
 \exp [ 
-(\Delta \phi -\Delta \phi_0 e^{-F\Omega_0 \tau } )^2/ 
2{\cal V}(\tau ) ] }{\sqrt{2\pi {\cal V}(\tau )}} , \,\,
        \label{F-P-sol1}  \\
&& \hspace{+0.6cm}
   {\cal V}(\tau) =(S_{\tilde\xi}/4F\Omega_0 ) \left(
1- e^{-2 F\Omega_0 \tau } \right) .
        \label{F-P-sol2}
        \end{eqnarray}
Calculating $\langle \cos \delta \phi (\tau )\rangle$ as 
$\int_{-\infty}^{\infty} \int_{-\infty}^{\infty} 
\cos [\Delta \phi -\Delta \phi_0] \linebreak[1]
\sigma (\Delta \phi,\tau |\Delta \phi_0)  \, 
 \sigma_{st}(\Delta \phi_0) \, d(\Delta \phi)\, 
d(\Delta \phi_0)$, we finally find the qubit correlation function
        \begin{equation}
K_z (\tau ) = \frac{\cos \Omega_0 \tau }{2} \, \exp \left[\frac{\cal C}{16 F}
\left( e^{-F\Omega_0 \tau} -1 \right) \right] .
        \label{K_z-an}
        \end{equation}
The validity range of this result is the same as for Eq.\ (\ref{D-analyt})
(${\cal C}\alt 1$, $16F/{\cal C} \agt 1$); we have checked that in this 
range Eq.\ (\ref{K_z-an}) fits well the numerical Monte-Carlo results. 
Fourier transform $S_z(\omega )=2\int_{-\infty}^{\infty} K_z(\tau )
e^{i\omega\tau} d\tau$ of Eq.\ (\ref{K_z-an}) in the case of efficient
feedback (${\cal C}/16F \alt 1$, so the exponent is expanded up to the 
linear term) gives the oscillation 
spectrum ($\omega >0$) 
        \begin{eqnarray} 
&& \hspace{-0.5cm} 
S_z(\omega )=\frac{1}{2}\left( 1-\frac{{\cal C}}{16F}\right) \, 
\delta \left( \frac{\omega -\Omega_0}{2\pi} \right) 
\nonumber \\ 
&& +\frac{{\cal C}}{8\Omega_0}\, \frac{1+F^2+(\omega /\Omega_0)^2}
{[1+F^2-(\omega /\Omega_0)^2]^2+4F^2(\omega/\Omega_0)^2} ,
        \label{S_z-an}
        \end{eqnarray} 
in which the first term ($\delta$-function) corresponds to synchronized 
non-decaying oscillations, while the second term describes fluctuations
and for $F\ll 1$  is peak-like near $\omega \approx \Omega_0$ with
the peak height of ${\cal C}/16\Omega_0 F^2$ and half-width at half-height
of $F\Omega_0$. 
[It is easy to check that 
$\int_0^\infty S_z (\omega ) \, d\omega /2\pi = 1/2$.] 

        Let us also calculate the correlation function 
of the detector current 
$K_I(\tau )=\langle [I(t+\tau )-I_0] [I(t) -I_0] \rangle$. 
Following Ref.\ \onlinecite{Kor-sp}, we use Eq.\ (\ref{I(t)}) to 
get $K_I(\tau )=(\Delta I/2)^2 K_z(\tau ) + K_{\xi}(\tau ) + 
(\Delta I/2)K_{z\xi}(\tau )$, where $K_{\xi}= (S_I/2)\, \delta (\tau)$ is 
due to pure noise while the cross-correlation term $K_{z\xi}(\tau )$ is due 
to quantum back-action, which shifts the phase $\phi$ by 
$-\sin\phi (\Delta I/S_I)\,\xi(t)\, dt$ as a result of noise $\xi$ acting 
during infinitesimal time $dt$ [see Eq.\ (\ref{ddphi})]. Because of the 
feedback, the effect of this extra phase shift decreases (on average) 
with time as 
$\tilde \delta \phi (\tau ) = - \exp (-F\Omega_0\tau)
[\sin\phi (t) ] (\Delta I/S_I)\, \xi(t)\, dt$ [see Eq.\ (\ref{F-P-sol1})] 
and the cross-correlation at $\tau >0$ can be calculated as
$K_{z\xi}(\tau )=\langle z(t+\tau)\, \xi(t)\rangle =
\langle \cos [\phi (t) +\Omega_0 \tau +\delta \phi (\tau )+ 
\tilde\delta \phi (\tau )] \, \xi (t) \rangle$. Expanding cosine 
up to the linear term in $\tilde\delta \phi (\tau )$ [the linear expansion
is the reason why it is sufficient to keep only averaged value 
$\tilde\delta \phi (\tau )$ instead of the full distribution], we obtain 
$K_{z\xi}(\tau ) = \langle \xi^2(t)\ dt\rangle (\Delta I/S_I)
\exp(-F\Omega_0\tau ) \langle 
\sin [\phi (t) +\Omega_0 \tau +\delta \phi (\tau )]\, \sin [\phi (t)] 
\rangle$,  where 
$\langle \xi^2 (t)\, dt \rangle =S_I/2$. Using symmetry of fluctuations 
leading to $\langle \sin \delta \phi (\tau ) \rangle =0$ (as above) 
and averaging over fast oscillations $\langle \sin [\phi (t)+\Omega_0 \tau ]
\sin [\phi (t)]\rangle = (\cos \Omega_0\tau )/2$, we finally obtain 
        \begin{equation}
K_{z\xi}(\tau) = \frac{\Delta I}{4} (\cos \Omega_0\tau ) \, 
e^{-F\Omega_0 \tau} 
\langle \cos \delta \phi (\tau )\rangle .
        \end{equation}
Since expression for $K_z(\tau )$ has a similar structure [see Eq.\ 
(\ref{K_z-via-dphi})], the corresponding terms of $K_I(\tau )$ 
are combined to yield 
        \begin{equation}
K_I(\tau )=  \frac{S_I}{2}\, \delta (\tau) +\frac{(\Delta I)^2}{4}
\left( 1+e^{-F\Omega_0\tau} \right) K_z(\tau ) ,
        \label{K_I}\end{equation}
where $K_z(\tau )$ is given by Eq.\ (\ref{K_z-an}).

 To calculate the 
spectral density $S_I(\omega )$ of the detector current, we again 
expand the outer exponent of Eq.\ (\ref{K_z-an}) up to the linear
term [validity of Eq.\ (\ref{K_I}) requires $16F/{\cal C}\agt 1$]; 
then the Fourier transform gives
        \begin{eqnarray}
&& \hspace{-0.9cm} S_I(\omega ) = S_I + \frac{(\Delta I)^2}{8} 
\left( 1-\frac{{\cal C}}{16F} \right) 
\delta \left(\frac{\omega-\Omega_0}{2\pi} \right)
     \nonumber \\
&& \hspace{-0.7cm} 
+\frac{S_I}{4} \frac{{\cal C}}{F} \frac{F^2[1+F^2+(\omega /\Omega_0)^2]}
{[1+F^2-(\omega /\Omega_0)^2]^2+4F^2(\omega/\Omega_0)^2} + {\mbox T4} ,
        \label{S_I} 
        \end{eqnarray}
where the last (fourth) term ${\mbox T4}$ is the same as the previous (third) 
term but with $F$ replaced by $2F$ and with extra factor ${\cal C}/16F$
[actually, higher-order terms of the exponent expansion will lead to 
extra terms with $F$ replaced by $3F$, $4F$, etc., 
and will slightly change the coefficients of the existing terms].
Notice that the $\delta$-function in the second term of Eq.\ (\ref{S_I}) 
is due to synchronized nondecaying oscillations, while the
third term at $F\ll 1$ describes a peak with height $(S_I/8)({\cal C}/F)$ 
and half-width $F\Omega_0$ near $\omega \approx \Omega_0$. 

It is easy to check that the integral
over all terms in Eq.\ (\ref{S_I}) except pure noise $S_I$, gives the 
total variance of the detector current equal to $(\Delta I)^2/4$
[this also follows directly from Eq.\ (\ref{K_I})], the same value as 
without the feedback.\cite{Kor-sp} Similarly to 
the non-feedback case, this variance would naively correspond to the 
qubit jumping between the two localized states, instead of oscillating
continuously [which would give twice smaller variance $(\Delta I)^2/8$]; 
in the Bayesian formalism this fact is understood as a consequence 
of non-classical cross-correlation between output noise and qubit evolution.

        Concluding this Section, let us emphasize again that in the ideal 
case the sufficiently strong feedback ($16F/{\cal C}\gg 1$) forces the qubit
evolution to be arbitrarily close to the perfect coherent oscillations 
running for arbitrarily long time. In this case the feedback efficiency 
$D$ approaches 
100\%, qubit correlation function becomes $K_z(\tau )=(\cos \Omega_0\tau )/2$,
in-phase quadrature component of the detector current becomes equal 
$(\Delta I)/4$, and the current spectral density contains (besides the
pure noise) the $\delta$-function peak at desired frequency $\Omega_0$
with variance $(\Delta I)^2/8$, and also the narrow peak around $\Omega_0$ 
(if ${\cal C}/16 \ll F \ll 1$) corresponding to same variance 
$(\Delta I)^2/8$.

\section{Effect of imperfect detector and extra dephasing} 

Various nonidealities reduce the fidelity of the quantum feedback
preventing $D$ from approaching 100\%. 
In this Section we consider the effects of imperfect quantum efficiency
of the detector ($\eta <1$) and extra qubit dephasing with rate 
$\gamma_{env}$ due to coupling to environment (see Fig.\ \ref{schematic}). 
Both effects contribute to the total qubit dephasing rate
$\gamma =\gamma_{env} + (\eta^{-1}-1)(\Delta I)^2/4S_I$ in Eq.\ (\ref{Bayes2})
and can be characterized by  effective quantum efficiency of
the qubit detection $\eta_{e}=[1+4\gamma S_I/(\Delta I)^2]^{-1} 
=[\eta^{-1}+4\gamma_{env} S_I/(\Delta I)^2]^{-1}$ or by effective relative 
dephasing $d_e= \gamma / [(\Delta I)^2/4S_I] = \eta^{-1} -1 + 
4\gamma_{env} S_I/(\Delta I)^2 = \eta_e^{-1}-1$; 
the physical meaning of $d_e$ is the ratio
of qubit coupling to sources of pure (unrecoverable) dephasing and qubit 
coupling to the detector governed by the quantum (informational) back-action.

\begin{figure}
\centering
\includegraphics[width=2.8in]{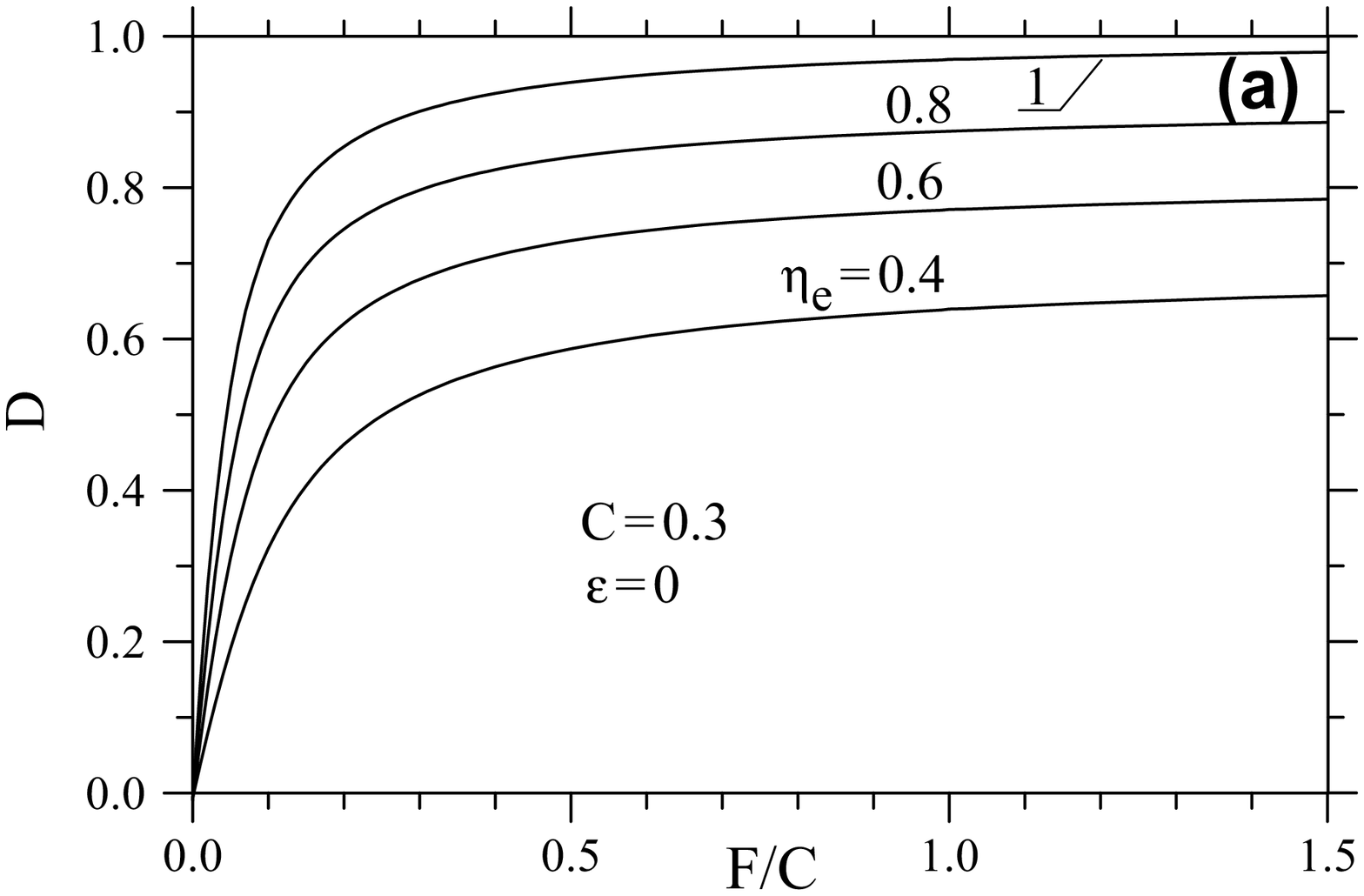}
\includegraphics[width=2.8in]{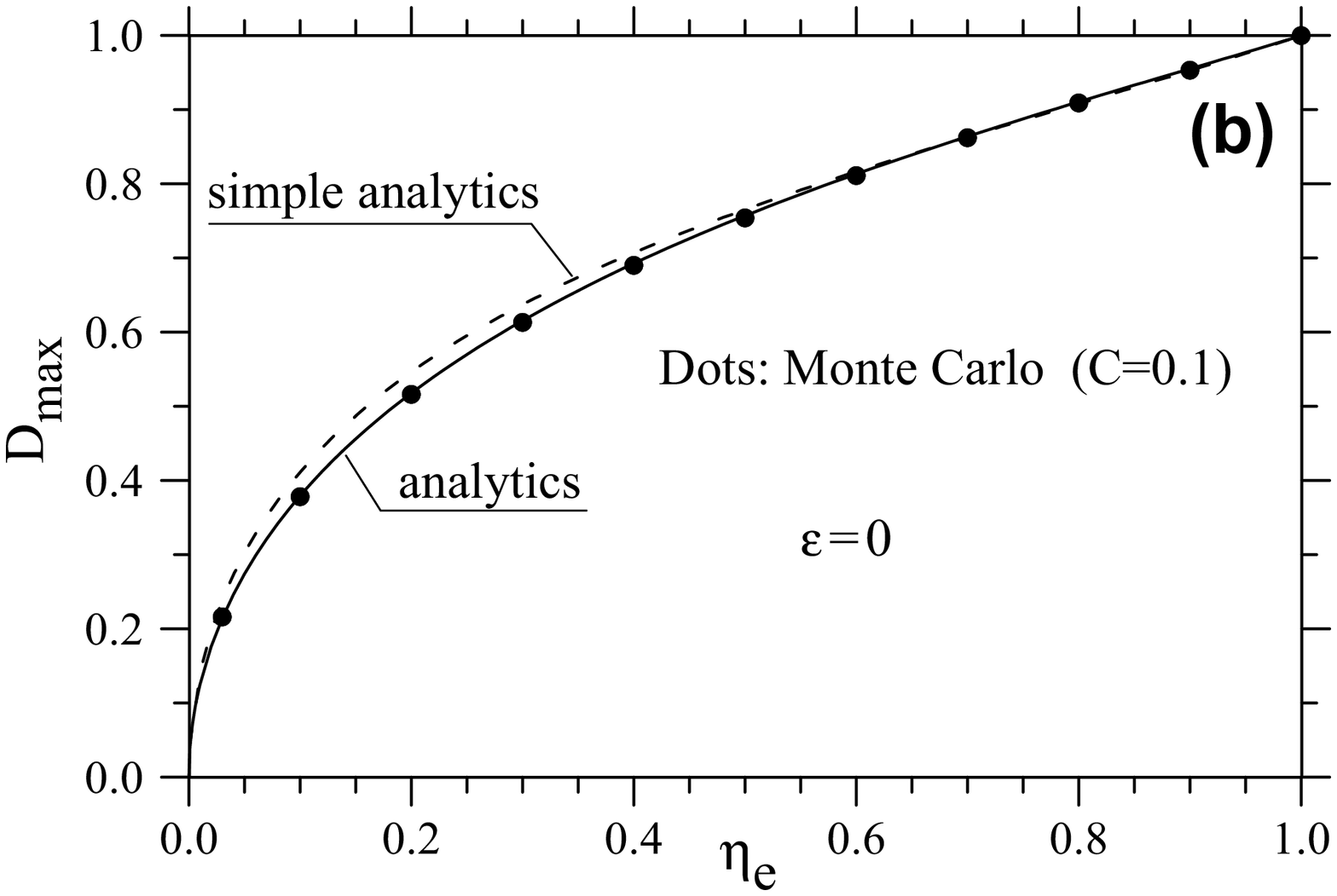}
%\vspace{0.4cm}
\caption{(a): Quantum feedback efficiency $D$ as a function of feedback 
strength $F$ 
for several values of quantum efficiency $\eta_e$ of the detection. 
(b): Maximum feedback efficiency $D_{max}$ (at large $F$) 
as a function of $\eta_e$. 
Dots show the Monte Carlo results for coupling ${\cal C}=0.1$, 
solid line corresponds to 
Eqs.\ (\ref{G(P2)})--(\ref{D-max-an}), and dashed line shows the approximate
formula (\ref{D-max-sim}).
 }
\label{fig-eta}\end{figure}

        Extra dephasing $d_e$ makes the qubit state non-pure; however, 
it is still perfectly monitored in a sense that $\rho^m (t)=\rho (t)$ 
(we assume that the magnitude of dephasing is known in the experiment
and is used in the processor solving the quantum Bayesian equations; 
we also still assume 
$\varepsilon =0$.) Therefore, controller (\ref{dHfb}) with sufficiently
large feedback factor $F$ can reduce the phase difference compared
to the desired oscillations practically to zero. As a result, we 
would expect that the feedback efficiency $D(F)$ should reach maximum
(saturate) at infinitely large $F$  similarly to the ideal case
shown in Fig.\ \ref{ideal}; however, this maximum will be less than unity. 
The saturating behavior of $D(F)$ dependence is confirmed by 
numerical (Monte Carlo) calculations -- see Fig.\ \ref{fig-eta}(a). 
Below we discuss the calculation of the saturated value 
$D_{max}$ at $F=\infty $ [Fig.\ \ref{fig-eta}(b)].

        The evolution of a non-pure qubit state with $\mbox{Re} \rho_{12}=0$
(since $\varepsilon =0$) can be parameterized as 
$\rho_{11}-\rho_{22} =P\cos \phi$, $\rho_{12}=iP(\sin \phi)/2$, where
purity factor $P$ is between 0 and 1. Using Bayesian equations 
(\ref{Bayes1})--(\ref{Bayes2}), we derive evolution equations for $P$
and $\phi$ (in Stratonovich form): 
        \begin{eqnarray}
&& \hspace{-0.9cm}
 \dot{P} = \frac{\Delta I}{S_I}  (1-P^2) 
( \frac{\Delta I}{2}\, P \cos \phi +\xi ) \cos \phi -\gamma P\sin^2 \phi ,
        \label{P-phi-evol-1}\\
&& \hspace{-0.9cm} 
\dot{\phi} = 2\frac{H_{fb}}{\hbar} -\frac{\sin\phi}{P} \frac{\Delta I}{S_I}
\, (\frac{\Delta I}{2}\, P \cos \phi +\xi ) -\gamma \frac{\sin 2\phi }{2} . 
        \label{P-phi-evol-2}
        \end{eqnarray}

Sufficiently strong feedback (\ref{dHfb}) makes the phase $\phi$ arbitrarily 
close to the desired phase $\phi =\Omega_0 t$ (mod $2\pi$), so the 
feedback efficiency is practically equal to the averaged purity factor: 
 $D_{max}=\langle P\rangle$. To find $\langle P\rangle$ in the case of 
weak coupling ${\cal C}/\eta_e \ll 1$,  let us perform first 
the averaging over oscillations and later the averaging over remaining slow 
fluctuations. It is easier to work with $P^2$ than with $P$, so we 
start with evolution equation for $P^2$ which is obtained from Eq.\ 
(\ref{P-phi-evol-1}) as $dP^2/dt = 2P\dot{P}$.
It is easier to average $P^2$ over oscillation period using the It\^o 
form \cite{Oksendal} because the noise $\xi$ causes correlated noise of 
$\phi$, and only in the It\^o form the average effect of the noise 
is zero. Using the standard rule \cite{Oksendal,Kor-01} we translate 
the evolution equation for $P^2$ into It\^o form:
        \begin{eqnarray}
&& \hspace{-0.5cm} 
\frac{dP^2}{dt} = \frac{(\Delta I)^2}{2S_I} (1-P^2)(1-P^2\cos^2\phi )
        -2\gamma P^2\, \sin^2 \phi 
        \nonumber \\
&& \hspace{0.5cm}
+(2\Delta I/S_I) P(1-P^2)(\cos \phi )\, \xi \, ;
        \end{eqnarray}
then averaging over $\phi$ is trivial: 
        \begin{equation}
\frac{dP^2}{dt} = \frac{(\Delta I)^2}{2S_I} (1-P^2)(1-\frac{P^2}{2} )
        -\gamma P^2 
+\frac{\sqrt{2}\Delta I}{S_I} P(1-P^2)\, \xi \, , 
        \label{dP2aver}
        \end{equation} 
where $\xi (t)$ is now a different white noise but with the same spectral 
density $S_\xi =S_I$, so we do not change notation.

        A simple estimate of $D_{max}$ can be obtained from Eq.\ 
(\ref{dP2aver}) by neglecting the noise term and finding stationary value 
for $P$, which gives\cite{Korotkov-04} 
        \begin{equation}
D_{max} \approx [1+1/2\eta_e -\sqrt{(1+1/2\eta_e)^2-2}]^{1/2}.
        \label{D-max-sim}
        \end{equation}
 
  If we do not neglect the noise term in Eq.\ (\ref{dP2aver}), then
$P^2$ fluctuates in time, and the stationary probability distribution 
$\sigma_{st}(P^2)$ can be found from the Fokker-Planck-Kolmogorov 
equation similar to Eq.\ (\ref{F-P}) (notice that varying diffusion 
coefficient comes inside the second derivative term) that leads
to equation 
        \begin{eqnarray}
&& 
[\gamma P^2 -(1-P^2)(1-P^2/2)(\Delta I)^2/2S_I] \sigma_{st}
\nonumber \\ 
&&
+[(\Delta I)^2/2S_I] \frac{d}{d(P^2)}[P^2 (1-P^2)^2\sigma_{st}] =0 ,
        \end{eqnarray}
which has analytic solution $\sigma_{st}(P^2) = N \, G(P^2)$,
where
        \begin{equation}
G(P^2) = (1-P^2)^{-5/2} \exp \left[ -\frac{\eta_e^{-1}-1}{2 (1-P^2)} 
\right]
        \label{G(P2)}
        \end{equation}
and $N$ is the normalization factor. Stationary probability distribution
for $P$ can be found as $\tilde{\sigma}_{st}(P)=2P\sigma_{st}(P^2)$, and 
calculating the average $P$ gives us finally the feedback efficiency 
        \begin{equation}
D_{max}= \frac{\int_0^1 P^2 \, G(P^2) \, dP}{\int_0^1 P \, G(P^2) \, dP} .
        \label{D-max-an}
        \end{equation}

        Figure \ref{fig-eta}(b) shows the dependence of the feedback 
efficiency $D_{max}$ on the effective quantum efficiency of the detector
$\eta_e=(1+d_e)^{-1}$. Solid line shows the analytical result 
(\ref{D-max-an}), dashed line shows approximate formula (\ref{D-max-sim}), 
and the symbols show the numerical (Monte Carlo) results for $D_{max}$
(at sufficiently large $F$) for coupling 
${\cal C}=0.1$. Notice that the lines for exact and approximate formulas 
are quite close to each other. 

        Since at finite detection efficiency $\eta_e$ the ensemble qubit 
dephasing 
is $\Gamma = \eta_e^{-1} (\Delta I)^2 /4S_I$, the weak coupling condition
requires ${\cal C}/\eta_e \alt 1$. As a result, the numerical results
for ${\cal C}=0.1$ in Fig.\ \ref{fig-eta} start to deviate (upwards) from the
analytical result at $\eta_e \alt 0.03$. For larger ${\cal C}$ the deviation 
starts even at larger $\eta_e$. 
Numerical calculations also show that at  ${\cal C}/ \eta_e \agt 3$ 
the average purity factor $\langle P\rangle $ has a noticeable dependence on 
the feedback factor $F$ ($\langle P\rangle $ decreases with increase of $F$),
while at  ${\cal C}/\eta_e \alt 1$ this dependence is negligible.

It is easy to see that in vicinity
of the ideal case ($\eta_e \approx 1$) Eq.\ (\ref{D-max-sim}) 
gives linear approximation $D_{max} \approx (1 +\eta_e)/2 \approx 1-d_e/2$ 
[exact solution (\ref{D-max-an}) shows the same linear approximation]. This 
explains the corresponding numerical result of Ref.\ \onlinecite{Ruskov-02}.
In the opposite limiting case $\eta_e \ll 1$, Eq.\ (\ref{D-max-sim}) 
is reduced to $D_{max}\approx \sqrt{2\eta_e}$; the exact solution 
(\ref{D-max-an}) has a similar dependence but with slightly different 
prefactor: $D_{max}\approx 1.25 \sqrt{\eta_e}$. 
Because of the square root dependence, feedback efficiency is still 
significant even for large magnitude of qubit dephasing due to 
coupling with environment. For example, if coupling with dephasing 
environment is 10 times stronger than coupling with quantum-limited 
detector ($d_e=10$, $\eta_e=1/11$), then $D_{max}\simeq 0.36$, which 
is still a quite significant value for an experiment.

\section{Effect of $\varepsilon$ and $H$ deviation} 

        In the ideal case we have assumed symmetric qubit ($\varepsilon=0$) 
and assumed that the exact value of tunneling parameter $H$ is 
used in the processor. In this Section we analyze what happens if
the qubit parameters $\varepsilon$ and $H$ deviate from the ``nominal'' 
values $\varepsilon =0$ 
and $H=H_0$ assumed by an experimentalist and used in the processor. 
In this case the monitored value $\rho^m$ of the qubit density matrix 
differs from the actual value $\rho$; and because of the mistake in qubit 
monitoring, the feedback performance should obviously worsen. 
[Both $\rho (t)$ and $\rho^m(t)$ satisfy Eqs.\ (\ref{Bayes1})--(\ref{Bayes2})
with the same detector output $I(t)$; however, ``incorrect'' parameters 
$\varepsilon =0$ and $H_0$ are used to calculate $\rho^m (t)$, while 
actual evolution $\rho (t)$ is governed by actual parameter values 
$\varepsilon$ and $H$.]
The desired evolution is still $\rho_{11}^d=(1+\cos \Omega_0 t)/2$,
$\rho_{12}^d=i(\sin \Omega_0 t)/2$ with $\Omega_0=2H_0/\hbar$,
which is used in calculation of feedback efficiency $D$
[notice that in the definition of efficiency (\ref{D-def}) $\rho^d (t)$ 
is multiplied by the actual density matrix $\rho (t)$, not the monitored 
value $\rho^m (t)$]. 
The controller is still given by Eq.\ (\ref{dHfb}) (we do not replace here
$H$ by $H_0$ because this is more natural, for example, for control 
of the Cooper-pair-box qubit). 
Since the analytical analysis of the problem is quite complicated, 
in this Section we present only the numerical results of Monte Carlo 
simulations.

\begin{figure}
\centering
\includegraphics[width=2.8in]{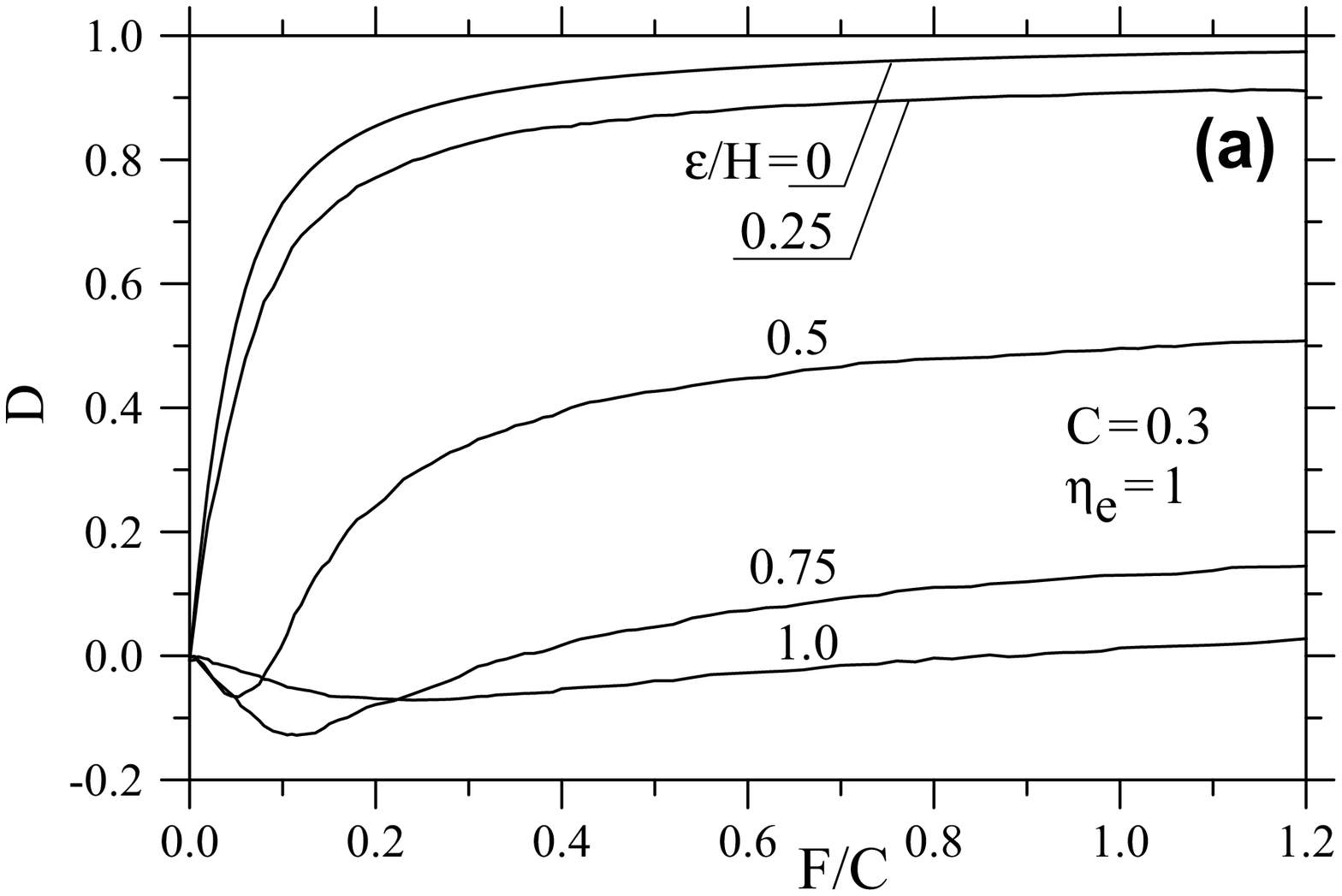}
\includegraphics[width=2.8in]{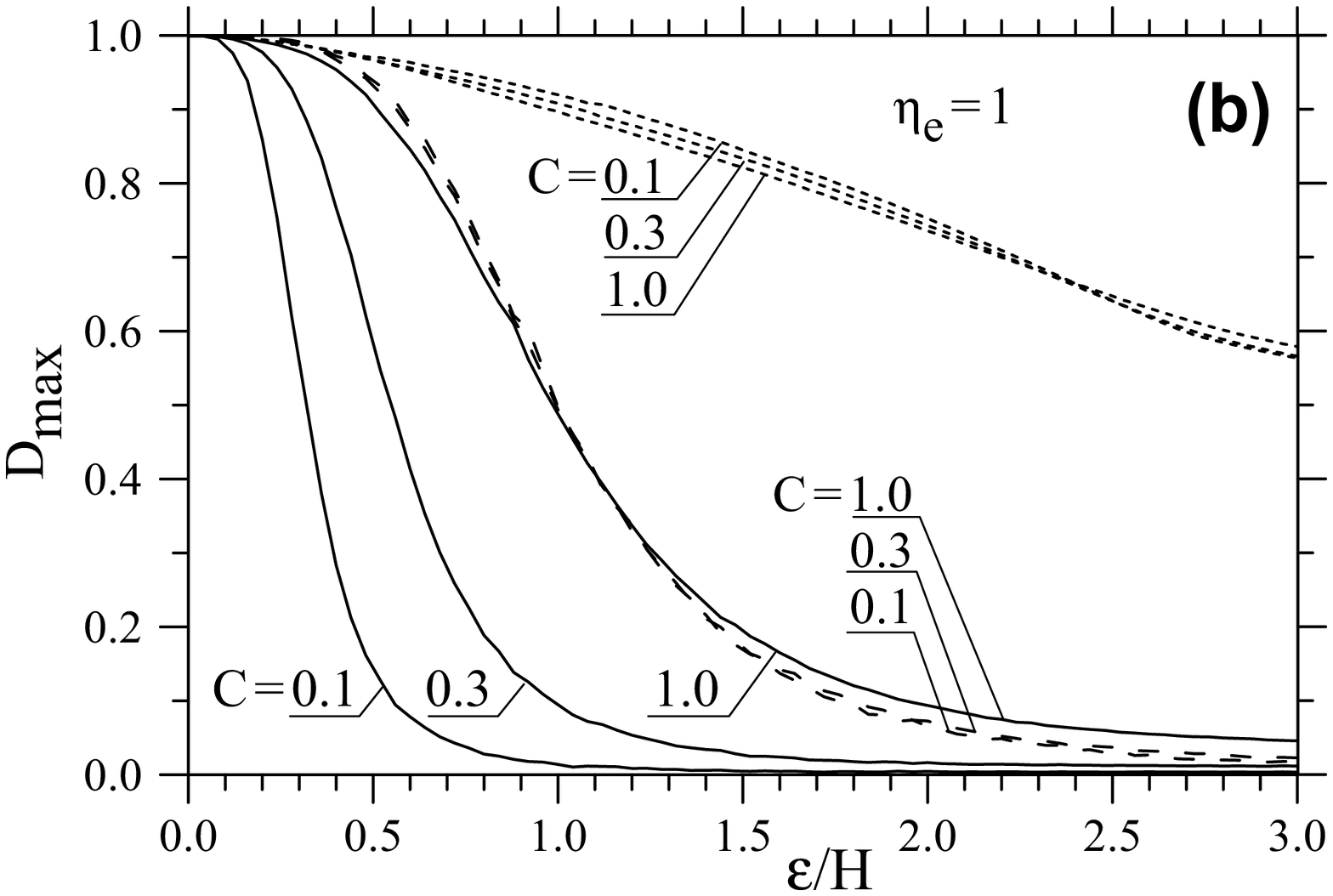}
%\vspace{0.4cm}
\caption{ (a): Dependence $D(F)$ for several values of the qubit energy 
asymmetry $\varepsilon /H$ in the case when the processor and controller 
still assume $\varepsilon =0$. (b): Solid lines: maximized over $F$ feedback
efficiency $D_{max}$ as a function of the asymmetry $\varepsilon /H$ 
for three values of coupling ${\cal C}=1$, 0.3, and 0.1. 
Dashed lines: the same curves for ${\cal C}=0.3$ and 0.1 drawn as functions
of $\varepsilon /H\sqrt{\cal C}$. Dotted lines: dependence 
$D_{max}(\varepsilon /H)$ for the three values of $\cal C$ 
in the case when actual value of $\varepsilon$ 
is used in the processor, while the controller (\ref{dHfb}) is still
designed for $\varepsilon =0$. 
 }
\label{fig-epsilon}\end{figure} 

        Let us start with deviation of $\varepsilon$ (while $H=H_0$). 
Figure \ref{fig-epsilon}(a) shows dependence $D(F)$ for ${\cal C}=0.3$ 
and several values of $\varepsilon$. 
One can see that for sufficiently large energy asymmetry $\varepsilon / H$ 
the feedback efficiency $D$ is negative at small $F$, while it is always 
positive at large $F$.  For relatively small values of asymmetry 
($|\varepsilon /H |\alt 1$) the dependence $D(F)$ apparently saturates 
at large $F$, while at larger asymmetry [$|\varepsilon /H |\agt 1.5$; not
shown in Fig.\ \ref{fig-epsilon}(a)]  
$D(F)$ has maximum at finite $F$. [We cannot exclude the possibility
that even for small $\varepsilon /H$, $D(F)$ also has maximum, but it occurs
at too large $F$ which cannot be analyzed by our code due to numerical 
problems.] Notice that the feedback efficiency is obviously insensitive to
the sign of energy asymmetry: 
$D(-\varepsilon, H,{\cal C}, F)= D(\varepsilon, H,{\cal C}, F)$.

        Solid lines in Fig.\ \ref{fig-epsilon}(b) show dependence 
of $D$ maximized over $F$, on energy asymmetry 
$\varepsilon /H$ for several values of the coupling ${\cal C}=0.1$, 
0.3, and 1. One can see that at small $\varepsilon /H$ the dependence 
$D_{max}(\varepsilon /H )$ is parabolic (zero derivative at $\varepsilon =0$),
which means that a small energy asymmetry of the qubit decreases the 
feedback efficiency very little. 
Zero derivative at $\varepsilon =0$ is a natural consequence of the symmetry
$D_{max}(-\varepsilon /H)=D_{max}(\varepsilon /H)$ (because of this symmetry,
we show only positive $\varepsilon /H$). 
As seen in Fig.\ \ref{fig-epsilon}(b), 
significant decrease of $D_{max}$ starts at smaller $\varepsilon /H$ 
for smaller coupling ${\cal C}$. Rescaling of the horizontal axis
by $\sqrt{\cal C}$ makes the curves quite close to each other (see dashed 
lines in the Figure); however, we are not sure if the scaling 
$D_{max}(\varepsilon /H \sqrt{\cal C})$ is 
really exact at ${\cal C}\rightarrow 0$. 

        The dotted lines in Fig.\ \ref{fig-epsilon}(b) show 
dependence $D_{max}(\varepsilon /H)$ for a different situation, 
when the exact 
value of $\varepsilon$ is used in the processor, but the controller
is still given by Eq.\ (\ref{dHfb}) designed for $\varepsilon =0$
[desired evolution is still given by Eq.\ (\ref{rho-desired}) with
$\Omega_0=2H/\hbar$].
 One can see that 
exact monitoring of the qubit significantly improves the feedback 
efficiency compared 
with the case considered above; however, the feedback efficiency still 
decreases with energy asymmetry because the desired evolution 
(\ref{rho-desired}) cannot be achieved at nonzero $\varepsilon /H$ and also
because of non-optimal controller still designed for $\varepsilon =0$. 
(Some apparent dependence of the dotted lines on ${\cal C}$ 
even at ${\cal C}\ll 1$ is possibly 
due to numerical problems of the code which does not work really well at
${\cal C} \alt 0.1$.)

\begin{figure}[t]
\centering
\includegraphics[width=2.8in]{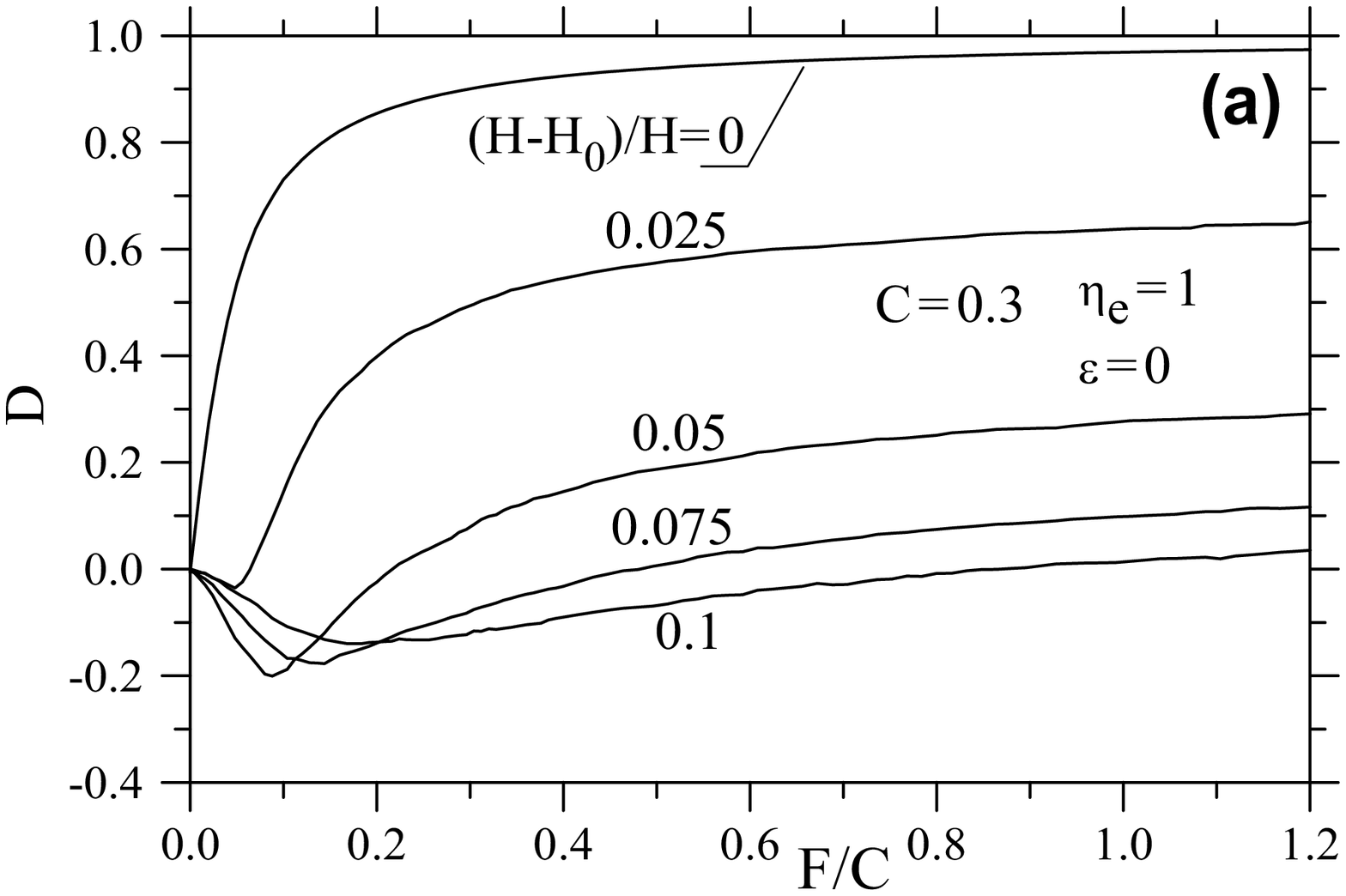}
\includegraphics[width=2.8in]{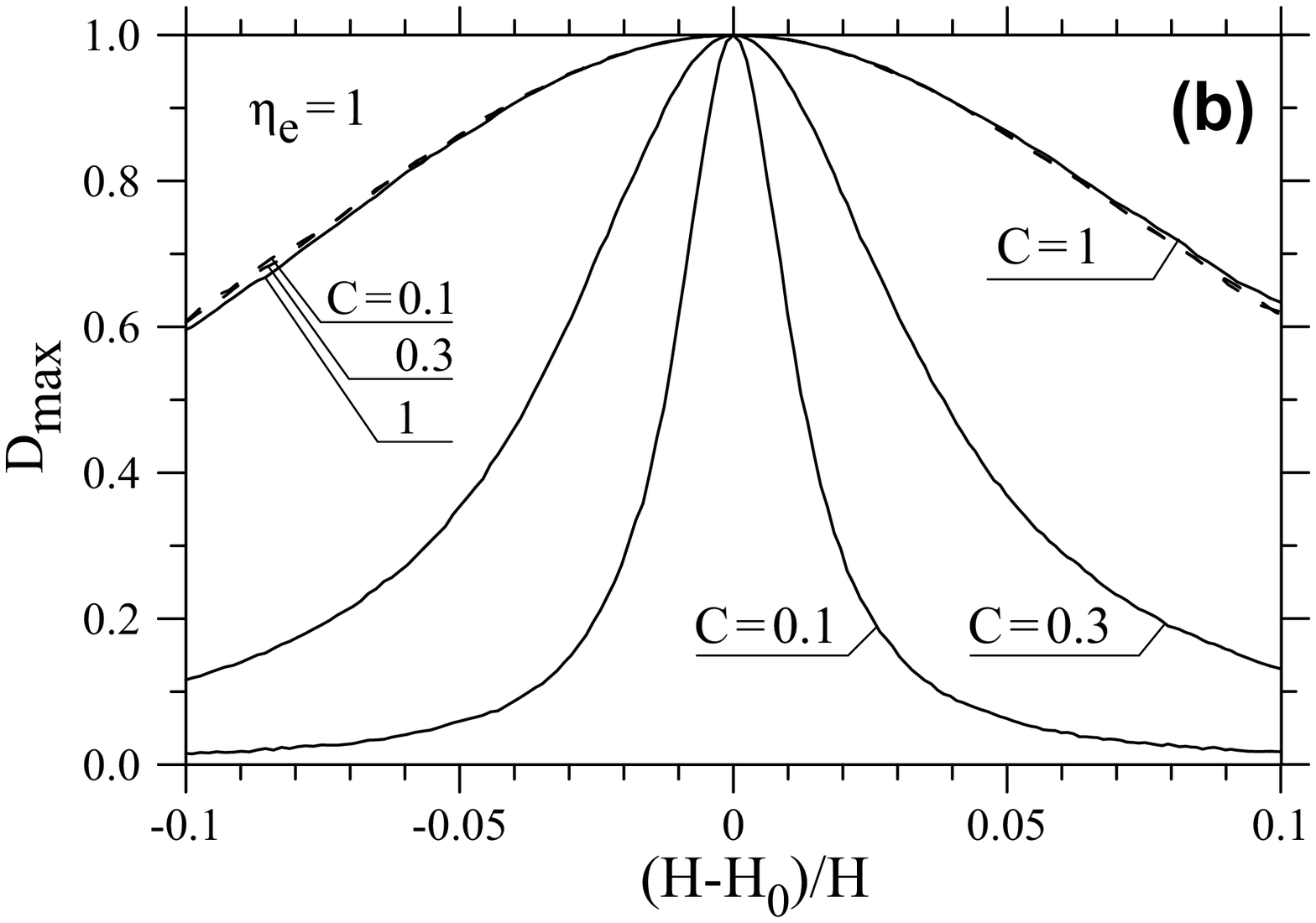}
%\vspace{0.4cm}
\caption{Effect of the deviation of the qubit parameter $H$ from the
value $H_0$ assumed in the processor. (a): Dependence $D(F)$ for
several values of the relative deviation $(H-H_0)/H$. (b): Solid lines: 
optimized over $F$ feedback efficiency $D_{max}$ as function of the
deviation $(H-H_0)/H$ for coupling ${\cal C}=1$, 0.3, and 0.1. Dashed lines:
the same curves for ${\cal C}=0.3$ and 0.1 drawn as functions of
$(H-H_0)/H{\cal C}$.
 }
\label{fig-H}\end{figure}

        To analyze the effect of the deviation of the parameter $H$
from the value $H_0$ used in the processor, we assume perfect energy 
symmetry, $\varepsilon =0$. Figure \ref{fig-H}(a) shows the dependence $D(F)$
for ${\cal C}=0.3$ and several values of the relative deviation $(H-H_0)/H$ 
(we show only the curves for positive deviation; the curves for negative
deviation are similar). 
We see that the effect of $H$ deviation is qualitatively similar to the
effect of energy asymmetry [compare Figs. \ref{fig-epsilon}(a) and
\ref{fig-H}(a)]. 
At large $F$ the dependence $D(F)$ saturates. 
Figure \ref{fig-H}(b) shows the value $D_{max}$ maximized over $F$ 
as a function of the relative deviation $(H-H_0)/H$ for several values of 
the coupling ${\cal C}$. One can see that the dependence is almost symmetric 
for positive and negative deviation, and is parabolic 
at small deviation similar to the case of nonzero $\varepsilon$ discussed
above. 
Also similar is the fact that weaker coupling $\cal C$ requires smaller 
deviation of $H$ for the same value of feedback efficiency. 
However, the scaling with ${\cal C}$ is now different: the curves become
close to each other if $D_{max}$ is plotted as a function of 
$(H-H_0)/H{\cal C}$ [see dashed lines in Fig.\ \ref{fig-H}(b)]. 
The different scaling is a natural consequence of the fact that 
small change of $\Omega=\sqrt{4H^2 +\varepsilon^2}/\hbar$ is linear in 
$H$ deviation but quadratic in $\varepsilon$. 
The results presented in Figs.\ \ref{fig-epsilon}(b) and \ref{fig-H}(b) 
can be crudely interpreted in the following way: $D_{max}$ decreases 
significantly when the Rabi frequency change due to parameter deviations
($\Delta \Omega = 2\Delta H/\hbar$ or $\Delta \Omega \approx 
\varepsilon^2/4H\hbar$) becomes comparable to the ``measurement rate''
$(\Delta I)^2/4S_I$. 
 [Notice that if the horizontal
axis in Fig.\ \ref{fig-H}(b) was chosen as $(H-H_0)/H_0$, the curves would 
be somewhat more asymmetric, the asymmetry being more significant at larger
$\cal C$.]

        Concluding the discussion of $\varepsilon$ and $H$ deviations,
let us mention that the main practical conclusion of the analysis is 
that the feedback operation is robust against small unknown deviations 
of the qubit parameters.

\section{Feedback control of a qubit with energy asymmetry $\varepsilon$}

    In this Section we analyze the case of a qubit with finite energy
asymmetry $\varepsilon$ (``asymmetric qubit''). 
In contrast to the problem considered in the 
previous Section, in which nonzero $\varepsilon$ was treated as an unwanted
deviation from the perfect zero value (therefore, finite $\varepsilon$ 
was just worsening the feedback designed for $\varepsilon =0$), now 
we try to design and analyze a different feedback (different controller) 
which goal is to maintain the free oscillations of a qubit with 
$\varepsilon \neq 0$ (so, now effect of nonzero $\varepsilon$ is 
what we also want to protect from decoherence). 
Hence, the desired evolution $\rho^d(t)$ is no longer given by Eq.\ 
(\ref{rho-desired}).

\begin{figure}
\centering
\includegraphics[width=2.0in]{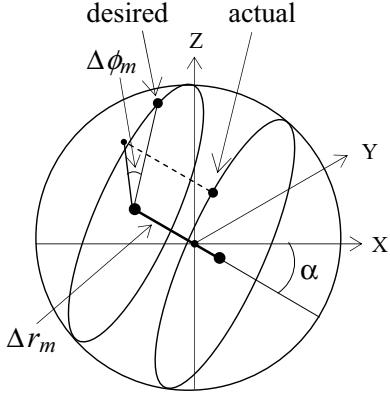}
%\vspace{0.4cm}
\caption{Illustration of the qubit evolution on the Bloch sphere. 
For an asymmetric qubit ($\varepsilon \neq 0$) the free evolution is
a rotation about a slanted (by angle $\alpha$) axis. The difference
between actual and desired qubit states (both are pure states) is
characterized by the distance $\Delta r_m$ between the corresponding
slanted planes and the angle $\Delta \phi_m$ within the slanted plane
(after projection). 
 }
\label{Bloch-sphere}\end{figure}

         Before choosing the desired evolution, let us mention
that the qubit asymmetry leads to one more degree of freedom on 
the Bloch sphere. In case of $\varepsilon =0$, a pure qubit state was 
characterized only by the phase $\phi$ [see Eq.\ (\ref{phi-def})] because
the real part of $\rho_{12}$ was vanishing in the course of measurement,
so the evolution was within the plane of ``zero longitude meridian''. 
For an asymmetric qubit, a naturally preferable plane of oscillations
on the Bloch sphere no longer exists; in particular a weak measurement 
leads to a slow fluctuation of the ``slanted'' plane of free qubit 
oscillations (Fig.\ \ref{Bloch-sphere}). The simulations show that 
without feedback the pure-state qubit evolution is to some 
extent confined between the two slanted planes passing through the 
``north pole'' ($\rho_{11}=1$) and ``south pole'' ($\rho_{22}=1$), 
with the probability about 0.6 of being between the two planes for
small $\cal C$ and $|\varepsilon /H| \alt 1$. 

        Let us choose the desired 
qubit evolution as a free evolution starting from the north pole:
    \begin{eqnarray}
&& \rho_{11}^d(t)= \frac{2H^2+\varepsilon^2 +2H^2\cos \Omega t}
{4H^2+\varepsilon^2}
\nonumber \\ 
&& \hspace{1cm}
= 1+\frac{1}{2}\cos ^2\alpha \,  (\cos \Omega t -1), 
        \label{desired-new1}  \\ 
&& \rho_{12}^d (t) = \frac{\varepsilon H (\cos \Omega t -1) }
{4H^2+\varepsilon^2} + \frac{iH\sin \Omega t}{(4H^2+\varepsilon^2)^{1/2}}
       \nonumber \\
&& \hspace{1cm}
 = \frac{\cos \alpha}{2} \, 
[\sin \alpha (\cos \Omega t -1) +i\sin \Omega t] .  
        \label{desired-new2} 
    \end{eqnarray} 
where $\Omega =\sqrt{4H^2+\varepsilon^2}/\hbar$ and 
$\alpha =\mbox{atan}(\varepsilon /2H)$. 
An interesting question is whether the quantum feedback can keep the qubit 
evolution close to the desired path 
(\ref{desired-new1})--(\ref{desired-new2}) or not.

        The old controller (\ref{dHfb}) is obviously not good for this 
purpose, so we need to design a new one. [Notice that the qubit density matrix
is monitored exactly, $\rho^m (t)=\rho (t)$, because all the parameters are 
assumed to be known exactly and because as discussed in Section II 
we assume infinite signal bandwidth.] As the first step, we characterize
the deviation of the monitored qubit state 
$\rho^m (t)$ from the desired state $\rho^d (t)$  by two 
magnitudes (see Fig.\ \ref{Bloch-sphere}): by the distance $\Delta r_m$ 
between two parallel slanted planes containing the monitored and desired 
states (the planes are slanted by  angle $-\alpha$) and 
by the angular difference $\Delta \phi_m$ between points $\rho^m$ and
$\rho^d$ projected onto the slanted plane. The corresponding formulas are 
a little lengthy but straightforward. For the distance deviation 
$\Delta r_m = r_m - r_d$ we calculate the distances $r_m$ and $r_d$ 
of the planes from the origin as the scalar products of the 
vector $(\cos \alpha,0,-\sin \alpha )$ orthogonal to the planes  
and the Bloch vectors 
$(2\mbox{Re}\rho_{12},2\mbox{Im}\rho_{12},\rho_{11}-\rho_{22})$ for 
the states $\rho^m$ and $\rho^d$, 
correspondingly: 
\begin{equation}
r_m = 2\mbox{Re}\rho_{12}^m  \cos \alpha - (\rho_{11}^m-\rho_{22}^m) 
\sin \alpha 
\end{equation}
and similar for $r_d$; 
it easy to see from Eqs.\ (\ref{desired-new1})--(\ref{desired-new2}) that 
$r_d=-\sin\alpha$.
 For the phase difference $\Delta \phi_m =\phi_m-\phi_d$ 
(mod $2\pi$, $|\Delta \phi_m|\leq \pi$) we use equation 
\begin{equation}
\mbox{tan}\,\phi_m = \frac{2\mbox{Im}\rho_{12}^m} 
{2\mbox{Re}\rho_{12}^m  \sin \alpha + (\rho_{11}^m-\rho_{22}^m) 
\cos \alpha } \, 
\end{equation}
[extra $\pi$-shift of $\phi_m$ is added when the denominator is negative 
as in Eq.\ (\ref{phi_m})]; 
a similar equation for $\rho^d$ defined by Eqs.\ 
(\ref{desired-new1})--(\ref{desired-new2})
obviously gives $\phi_d=\Omega t$ (mod $2\pi$). Notice that in the case 
$\varepsilon=0$ (so that $\alpha =0$) we recover the previous definition 
(\ref{phi_m}) of $\Delta \phi_m$, while $\Delta r_m =0$.

        Limiting ourselves by the feedback control of the qubit parameter 
$H$ only, we designed and analyzed the following controller: 
\begin{equation}
  \Delta H_{fb}= -F H \Delta \phi_m - F_r H \sin\phi_m \Delta r_m . 
        \label{new-controller} 
\end{equation}
The first term in this expression is the same as in the previous controller
(\ref{dHfb}) and is supposed to reduce the in-plane phase difference 
$\Delta \phi_m$ by changing the oscillation frequency, 
while the second term is supposed to reduce the
inter-plane distance $\Delta r_m$. The idea is that the change of 
$H_{fb}=H+\Delta H_{fb}$ affects the angle of the slanted plane of 
oscillations, and when it is done periodically in phase with the 
oscillations (due to the factor $\sin \phi_m$) the inter-plane distance
can be gradually reduced. 

\begin{figure}
\centering
\includegraphics[width=3.1in]{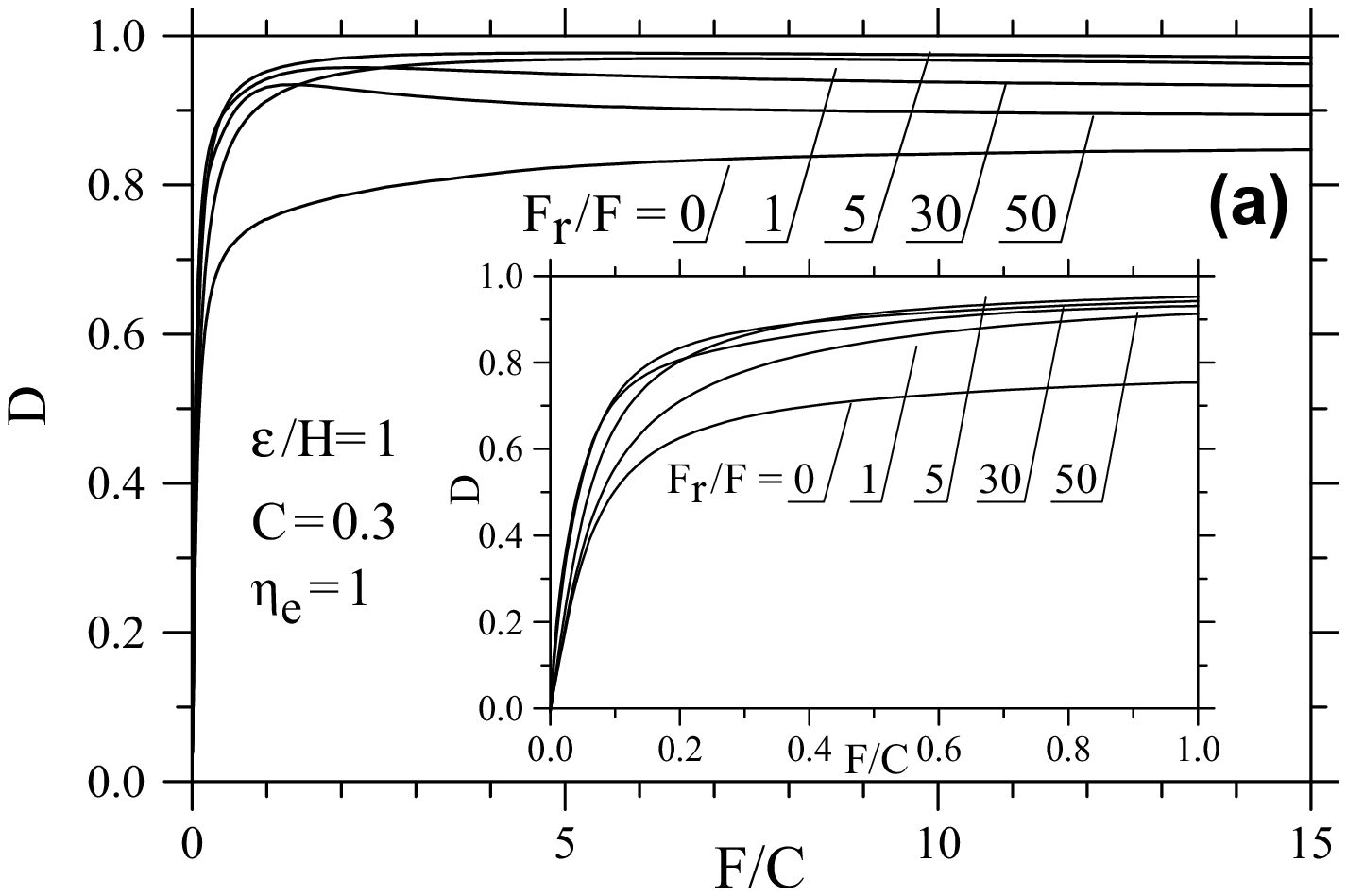} 
%\vspace{0.2cm}
\includegraphics[width=2.8in]{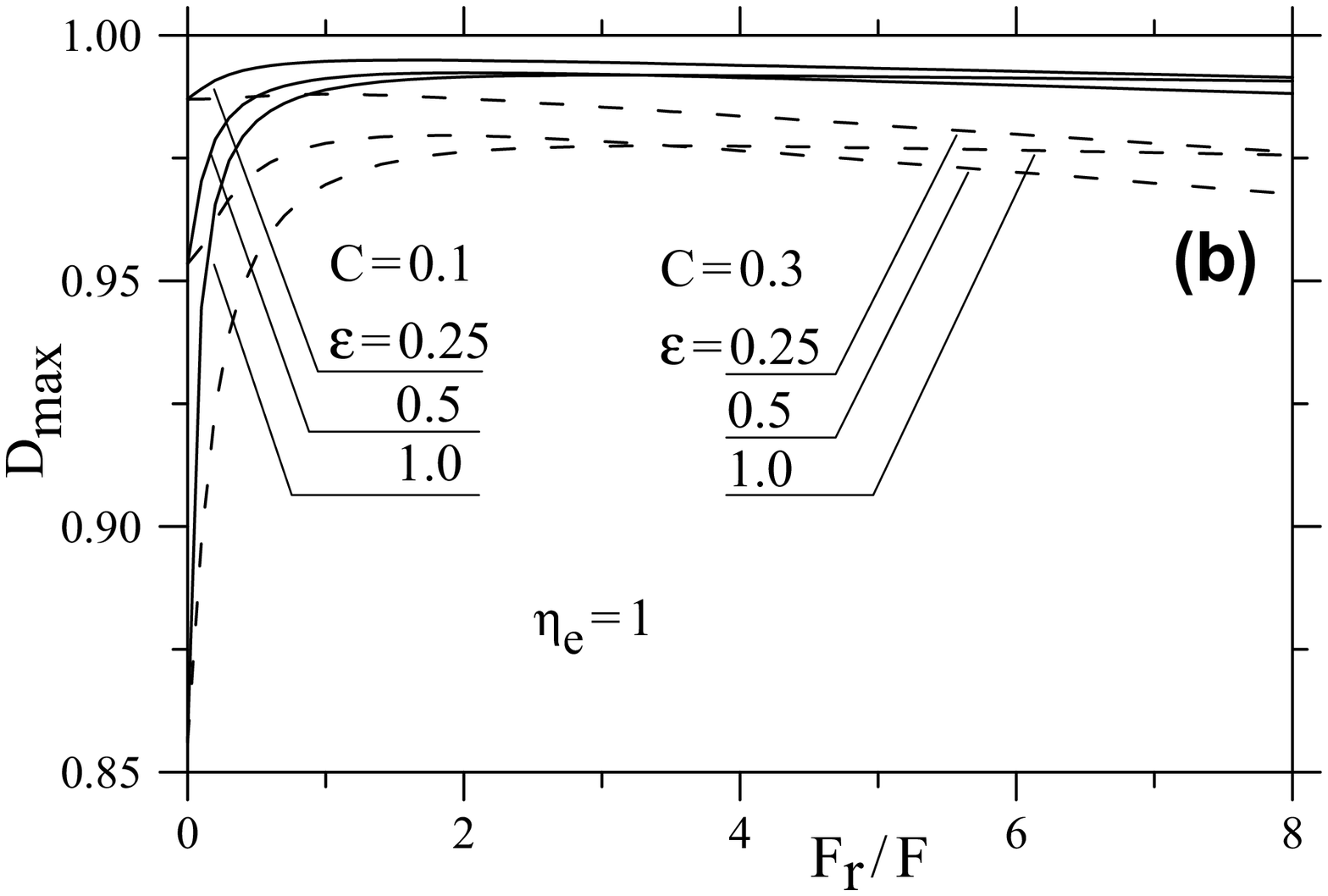} 
\includegraphics[width=2.8in]{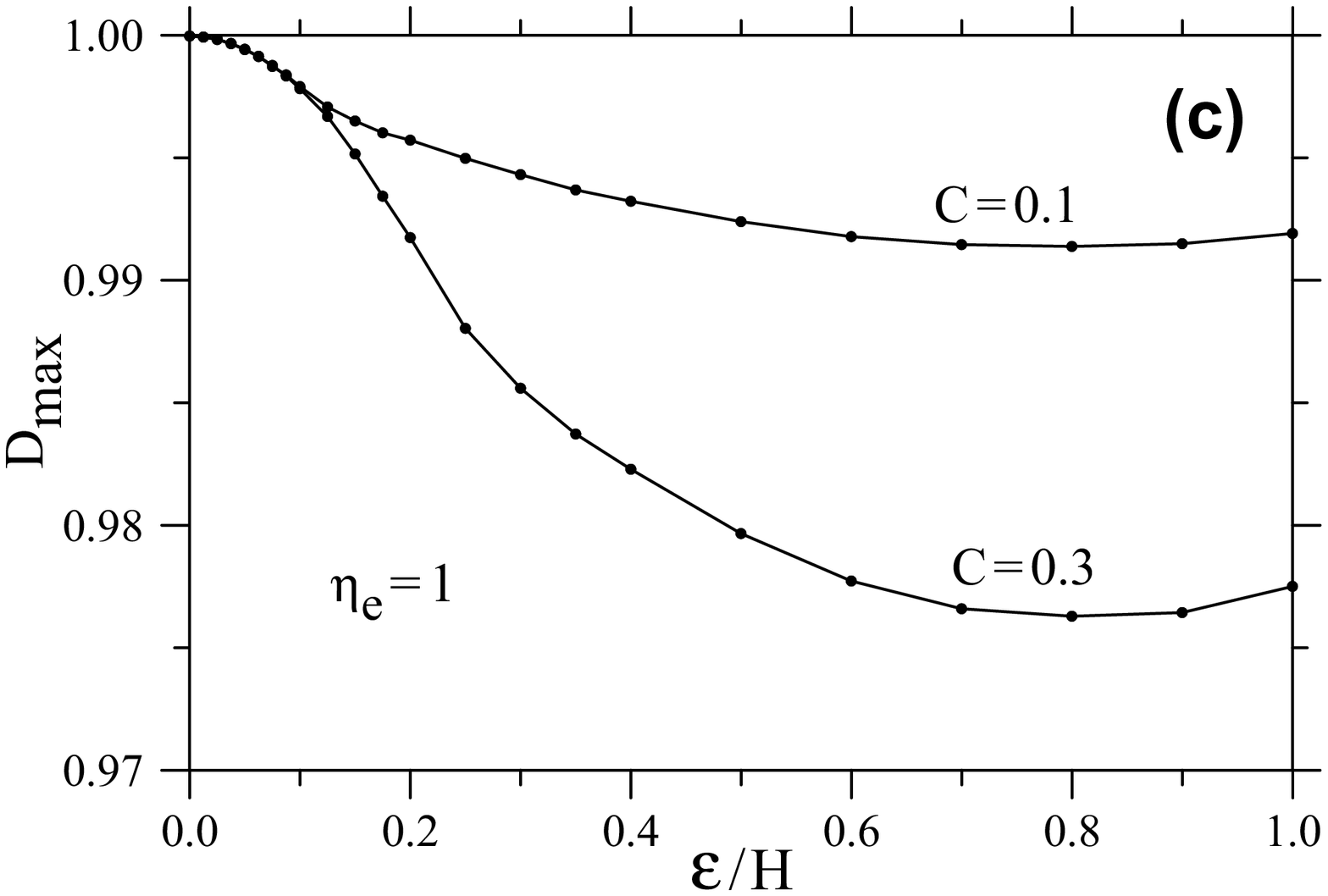} 
%\vspace{0.4cm}
\caption{Feedback efficiency for an energy-asymmetric qubit 
($\varepsilon\neq 0$). 
(a): Dependence $D(F)$ for several values of the ratio $F_r/F$. 
Inset shows the same curves at small $F$. 
(b): Optimized over $F$ feedback efficiency $D_{max}$ as a function of 
$F_r/F$ for several values of qubit asymmetry $\varepsilon /H$ and two 
values of coupling ${\cal C}$. 
(c): Feedback efficiency $D_{max}$ optimized over both $F$ and $F_r$, 
as a function of asymmetry $\varepsilon /H$ for two values of ${\cal C}$.
Dots show numerical results while the lines just connect the dots. 
 }
\label{fig-eps-1}\end{figure} 

        Numerical calculations show that this idea works really well. 
Figure \ref{fig-eps-1}(a) shows the dependence $D(F)$ for several values 
of the ratio $F_r/F$ using as an example parameters $\varepsilon /H= 1$, 
${\cal C}=0.3$, and $\eta_e=1$. One can see that non-zero $F_r$ 
can significantly improve the feedback efficiency $D$. While at $F_r=0$
the dependence $D(F)$ saturates at large $F$, at non-zero $F_r$ the 
efficiency $D$ has maximum at finite $F$. 

Figure \ref{fig-eps-1}(b) shows the optimized over $F$ efficiency $D_{max}$
as function of the ratio $F_r/F$ for couplings ${\cal C}=0.3$ and 0.1,
and three values of energy asymmetry $\varepsilon /H$. One can see that
each curve has maximum at some value of $F_r/F$. Notice also that at
zero $F_r$, the curves for different coupling ${\cal C}$ practically coincide,
while at finite $F_r/F$ their behavior significantly depends on ${\cal C}$,
with larger $D_{max}$ at smaller coupling. 

In Fig. \ref{fig-eps-1}(c) we show the feedback efficiency $D_{max}$ 
optimized over both $F$ and $F_r$ as function of energy asymmetry 
$\varepsilon /H$ for two values of the coupling ${\cal C}$. As we see, 
finite asymmetry $\varepsilon /H$ prevents efficiency $D_{max}$ from
reaching 100\%. However, the difference $1-D_{max}$ decreases with decrease
of coupling ${\cal C}$, crudely proportional to ${\cal C}$ (except
the region of small $\varepsilon /H$, where the accuracy of our calculations
is possibly insufficient to distinguish the curves; unfortunately,
there is no simple way to estimate the calculation accuracy). 
Therefore, we guess 
that for any asymmetry $\varepsilon /H$, the feedback efficiency $D_{max}$ 
reaches 100\% in the limit of small coupling ${\cal C}\rightarrow 0$.
(We cannot check this conjecture numerically because our code does not work
well at ${\cal C}<0.1$.)

       Limiting the feedback control by the control of the parameter $H$
only, is not quite natural for the asymmetric qubit (though it is simpler 
from the experimental point of view). We have also performed a preliminary
analysis of a simultaneous control of both $H$ and $\varepsilon$. 
We have considered the case when Eq.\ (\ref{new-controller}) is used for 
$H$-feedback while a similar equation (with $H$ replaced by $\varepsilon$)
is used for simultaneous control of $\varepsilon$. Even though we have not
performed detailed optimization, we have obtained larger values of $D_{max}$
than those presented in Figs.\ \ref{fig-eps-1}(b) and \ref{fig-eps-1}(c).
This shows that additional feedback control of the qubit parameter 
$\varepsilon$ really improves the feedback efficiency.

    Concluding this Section let us mention that its main result is the 
possibility of a very efficient feedback control of an asymmetric qubit. 
This can be done even using the control of the parameter $H$ only, 
while simultaneous control of $\varepsilon$ further improves the operation 
of the feedback.

\section{Conclusion}

    In this paper we have analyzed the quantum feedback control of a single 
qubit, designed  to maintain perfect (or close to perfect) 
Rabi oscillations for arbitrarily long time. We have considered ``Bayesian''
feedback \cite{Ruskov-02} which requires a ``processor'' (see Fig.\ 
\ref{schematic}) solving quantum Bayesian equations to monitor the qubit
state evolution via continuous output signal from the detector (QPC or SET) 
weakly coupled 
to the qubit. After comparing the randomly evolving (due to quantum 
back-action) monitored qubit state $\rho^m$ with the desired state $\rho^d$,
the qubit tunneling parameter $H$ is being slightly changed in order
to reduce the difference between the states. 
 For simplicity we have assumed infinite bandwidth of the (noisy) 
detector signal and neglected the time delay in the feedback loop. 

     The analysis in Section III shows that in the ideal case the efficiency 
$D$ of the quantum feedback can be made arbitrarily close 
to 100\% by increasing the strength of the feedback control [characterized
by parameter $F$ in Eq.\ (\ref{dHfb})]. It is important to mention that $F$ 
scales with the coupling ${\cal C}$ between qubit and detector; therefore
in the realistic case of weak coupling ${\cal C}\ll 1$, the parameter $F$
and consequently the relative change of the qubit parameter $H$ remain small.
The efficient operation of the feedback loop is achieved at $F/{\cal C} \gg 1$
[see Eq.\ (\ref{D-analyt})]; in this case the qubit evolution becomes
almost perfectly sinusoidal [Eqs.\ (\ref{K_z-an}) and (\ref{S_z-an})],
while the spectral density of the detector current [Eq.\ (\ref{S_I})]
contains the $\delta$-function peak at Rabi frequency $\Omega_0$ with the 
integral $(\Delta I)^2/8$ (as would be expected for the synchronized classical
sinusoidal oscillations in the qubit) and also a narrow peak around 
$\Omega_0$ with the same integral. The total integral under the peaks is thus
$(\Delta I)^2/4$, which exceeds the limit for a classically interpretable
process.\cite{Ruskov-05} 

    The feedback performance worsens in the case of a nonideal detector
and/or presence of dephasing environment. This case is considered in Section
IV. We have obtained an analytical formula [Eqs.\ 
(\ref{G(P2)})--(\ref{D-max-an})] for the maximum feedback efficiency 
$D_{max}$ confirmed by Monte Carlo simulations. 
It gives $D_{max}\approx (1+\eta_e)/2$ in almost perfect case when the 
effective detection efficiency $\eta_e$ is close to unity, and
$D_{max}\approx 1.25 \sqrt{\eta_e}$ when $\eta_e\ll 1$.

    In Section V we have analyzed numerically the decrease of the 
feedback efficiency in the case when actual qubit parameters $\varepsilon$
and $H$ differ from the assumed (in the processor and controller) parameters
$\varepsilon =0$ and $H=H_0$ (otherwise the case is ideal). 
We have found that for small deviations the efficiency $D_{max}$ decreases 
relatively slowly (with zero derivative at vanishing deviation),
so that, for example, $D_{max}\geq 0.95$ is possible for $|\varepsilon /H_0| 
< 0.5 \sqrt{\cal C}$ and $|H/H_0-1|<0.03\, {\cal C}$. This shows that
the quantum feedback is robust against small deviations of the qubit
parameters. 
  
   In Section VI we have analyzed the feedback control of a qubit with
finite energy asymmetry $\varepsilon$, so that the desired evolution 
trajectory is along a slanted circle on the Bloch sphere. Despite the 
control problem becomes two-dimensional in this case even for a pure state, 
we have shown
that efficient feedback is still possible using only one controlled parameter 
$H$ and properly designed algorithm (controller). 

    The Bayesian quantum feedback of a solid-state qubit analyzed in this 
paper is not yet realizable experimentally at the present-day level 
of technology. Even much simpler quadrature-based quantum feedback
\cite{Korotkov-04} is still a big experimental challenge. However, a rapid 
progress in experiments with solid-state qubit and also recent realization 
of quantum feedback in optics \cite{Geremia} allow us to believe that
the analysis performed in this paper will eventually be experimentally
relevant. 
In this paper we have not considered two more effects quite important for 
the operation of the Bayesian quantum feedback: finite signal bandwidth and 
time delay in the loop. These effects will be a subject of a separate 
publication. 

\vspace{0.3cm}

The work was supported by NSA and ARDA under ARO grant W911NF-04-1-0204.


\begin{thebibliography}{99}
\bibitem{H-sign} It may be more meaningful to define $H$ as a negative 
        real number to have a natural form for the qubit ground state;
        however, we prefer the choice of positive $H$ that can always be
        done by changing the phase of one of the localized states.
\end{thebibliography}
\end{document}